\documentclass[twocolumn,nolinenumbers,trackchanges]{aastex701}
\usepackage{amsmath}
\usepackage{amssymb}
\usepackage{amsthm}
\usepackage{natbib,aasdefs,url,bm}
\usepackage{array}
\usepackage{float}
\usepackage{graphicx}
\usepackage{subfigure}

\def\be{\begin{equation}}
\def\ee{\end{equation}}
\def\ba{\begin{eqnarray}}
\def\ea{\end{eqnarray}}

\def\aap{\ {A\&A}\ }

\def\apjl{\ {ApJL}\ }
\def\apjs{\ {ApJS}\ }

\submitjournal{ApJL}
\accepted{July 21, 2025}
\shorttitle{EM signatures from pulsar remnants of BNS mergers}
\shortauthors{Mukhopadhyay and Kimura}
\begin{document}
\title{Electromagnetic signatures from pulsar remnants of binary neutron star mergers: prospects for unique identification using multi-wavelength signatures}
\correspondingauthor{Mainak Mukhopadhyay}
\email{mkm7190@psu.edu}
\author[0000-0002-2109-5315]{Mainak Mukhopadhyay}
\affiliation{Department of Physics; Department of Astronomy \& Astrophysics; Center for Multimessenger Astrophysics, Institute for Gravitation and the Cosmos, The Pennsylvania State University, University Park, PA 16802, USA}
\email{mkm7190@psu.edu}
\author[0000-0003-2579-7266]{Shigeo S. Kimura}
\affiliation{Frontier Research Institute for Interdisciplinary Sciences; Astronomical Institute, Graduate School of Science, Tohoku University, Sendai 980-8578, Japan}
\email{shigeo.s.ti@gamil.com}
\begin{abstract}
Binary neutron star (BNS) mergers can result in the formation of long-lived magnetar remnants which can enhance neutrino and electromagnetic (EM) emissions. In this work, we study the resulting multi-wavelength EM emissions and the prospects of their detectability in the current and upcoming EM telescopes. We model the pulsar-wind neubla system where the long-lived pulsar with dipolar magnetic fields of $10^{13} - 10^{15}$ G (magnetar) spins down and is surrounded by an outward expanding nebula and kilonova ejecta. Although at early times post the merger, the EM signatures are unobservable due to heavy attenuation, they become observable on timescales of $\mathcal{O}(1 - 10)$ days after the merger. We find that the survey and follow-up observations have horizon distances $\gtrsim 1\ \rm Gpc$ for most of the wavebands and conclude that the detection prospects for such long-lived remnants in the electromagnetic channel are promising. This is of crucial importance for multi-messenger observations from BNS mergers to constrain the physical parameters of the remnants. Furthermore, we highlight how observations across the electromagnetic band can uniquely identify magnetar-powered transients resulting from BNS mergers and establish concrete associations of the detected gravitational wave signatures with such transients.
\end{abstract}
\section{Introduction}
\label{sec:intro}
Since the discovery of GW170817 in the gravitational wave (GW)~\citep{LIGOScientific:2017vwq} and electromagnetic (EM)~\citep{Goldstein:2017mmi,LIGOScientific:2017zic,LIGOScientific:2017ync,DES:2017kbs,Coulter:2017wya,J-GEM:2017tyx,Valenti:2017ngx,Lipunov:2017dwd,Chornock:2017sdf,Drout:2017ijr,Haggard:2017qne,Hallinan:2017woc,Kilpatrick:2017mhz,Margutti:2017cjl,Pian:2017gtc,Savchenko:2017ffs,Shappee:2017zly,Smartt:2017fuw,Troja:2017nqp,DAvanzo:2018zyz,Ghirlanda:2018uyx} channels, the multi-messenger paradigm associated with GWs and photons as a result of compact object mergers has been established. The remnant of GW170817 likely collapsed to a black hole given the mass of the system~\citep{Pooley:2017mzo,Shibata:2019ctb,Gill:2019bvq}, but there is some debate that the remnant could also be a neutron star~\citep{Piro:2018bpl,DuPont:2024sbz}. Possible EM signatures resulting from binary neutron star (BNS) mergers or from the remnants thus present an intriguing case for observational astronomy and insights for theoretical astrophysics.

The inconclusive evidence regarding the central remnant of GW170817 points toward a more generic issue, that the fate of a BNS merger is fairly uncertain. It mainly depends on the nuclear equation of state, mass, and spin of the component neutron stars, which are not well determined. Amongst the various outcomes possible~\citep{Sarin:2020gxb}, an interesting scenario is the formation of a long-lived (stable) differentially spinning neutron star with a millisecond period - a pulsar~\citep{Giacomazzo:2013uua,Radice:2018xqa,Shibata:2019wef,Sarin:2020gxb}. In some cases, these pulsars can have a large dipolar magnetic field $B_d \sim \mathcal{O}(10^{13}\ {\rm G} - 10^{15}\ {\rm G})$~\citep{Price:2006fi,Akgun:2013aq,Giacomazzo:2014qba,Kiuchi:2023obe}, which can then be termed as \emph{magnetars}. 

Such a magnetar-powered source leads to an enhanced luminosity in the EM channel at timescales comparable to the spindown timescale, typically ranging from days to weeks~\citep{Murase:2010fq,Bucciantini:2011kx,Murase:2014bfa,Murase:2017snw,Sarin:2022wby,Ren:2022tot,Omand:2024yye}. Previous works like~\cite{Omand:2024yye} and~\cite{Ren:2022tot}, have also explored broadband non-thermal emissions from pulsar-wind nebula, where the former investigated their implications for GRB170817A and GRB210702A, while the latter investigated AT 2017gfo (UV/optical/IR bands) and the late-time X-ray afterglow of GRB 170817A associated with GW 170817. However in the current work we do a more detailed and improved calculation that includes better opacity modeling across all wavelengths, computation of gamma-rays (which will prove to be crucial in distinguishing magnetar-powered transients from the rest) amongst other things. Furthermore, our scenario does not necessarily include a jet and thus the total spindown luminosity is injected into the nebula.

We explored the high-energy neutrino signatures from such a source in a previous work~\citep{Mukhopadhyay:2024ehs} (see also~\cite{Murase:2009pg,Gao:2013rxa,Fang:2013vla,Fang:2017tla,Fang:2018hjp}), where we also discussed the possibility of detecting neutrinos from such a source using GW signatures, implementing a triggered stacking search~\citep{Mukhopadhyay:2023niv,Mukhopadhyay:2024lwq}. The central question we want to answer in this \emph{Letter} is - what EM counterparts can be expected if the resulting remnant from a BNS merger forms a magnetar wind-nebulae-ejecta system? 

Several novel aspects are discussed and presented in this work. We explore non-thermal emission resulting in a multi-wavelength EM signal. Unlike in the case of gamma-ray bursts (GRBs), in our case the resulting spindown energy of the magnetar is isotropically distributed, enabling the detection of non-thermal EM counterparts independent of the inclination. We explicitly compute the redshift-dependent detection horizons for various current and upcoming EM telescopes across all wavelengths. Furthermore, we outline how multi-wavelength signatures from such long lived magnetar remnants resulting from BNS mergers can be \emph{uniquely} associated with the corresponding GW signatures. The implications of our study are to provide conclusive evidence regarding the central merger remnant. Furthermore, late time EM emission can help characterize the physical properties of the central engine, provide precise localization information for associated searches for high-energy neutrinos, and help with detecting and classifying sub-threshold GW events~\citep{Goldstein:2019tfz}. 

The \emph{Letter} is organized as follows. We present an overview of the magnetar-nebula-ejecta model in Section~\ref{sec:model}. The main results of this work along with the resulting photon spectra and light curves are discussed in Section~\ref{sec:res}. In Section~\ref{sec:det} we present the detection prospects for EM signatures including multi-wavelength characteristics to identify and disentangle them from other transients. We conclude in Section~\ref{sec:disc}. 
\section{Model Overview}
\label{sec:model}
In this section, we briefly overview the magnetar-nebula-ejecta model. The model is discussed in detail in~\cite{Mukhopadhyay:2024ehs} (see Figure 1 there for a schematic figure). Here we will just summarize the main aspects of the model that would be relevant for the EM signatures.
\begin{figure}
\centering
\includegraphics[width=0.48\textwidth]{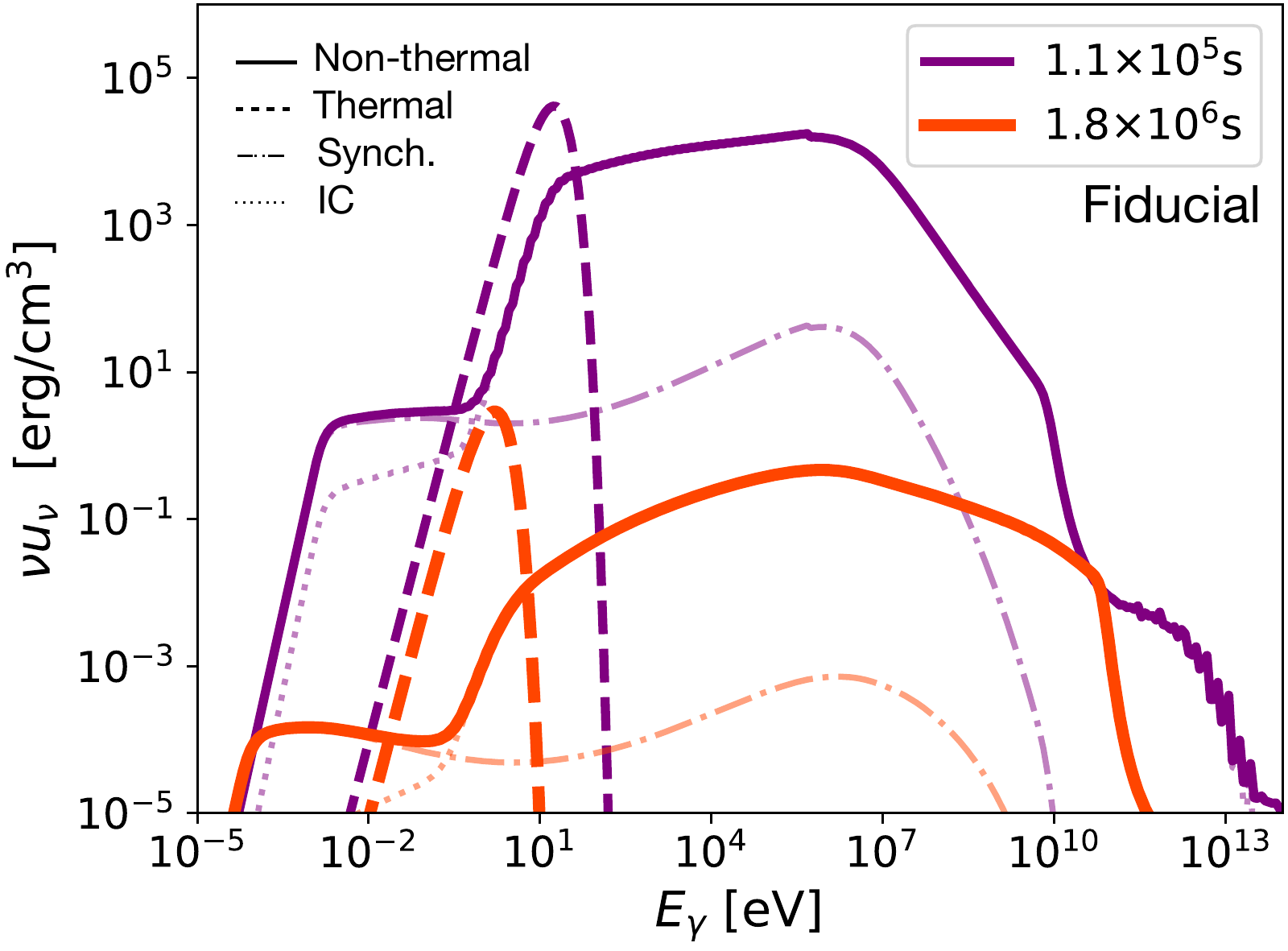}
\caption{\label{fig:photonspectra}Photon energy density spectrum ($\nu u_\nu$) of the non-thermal (solid) and thermal (dashed) photons in the nebula (comoving frame) after $\gamma \gamma$ and SSA attenuations, at different times for the fiducial case. The corresponding synchrotron (Synch.) and inverse Compton (IC) components are shown with thin dashed-dot and dotted lines respectively.
}
\end{figure}
\begin{figure*}
\centering
\includegraphics[width=0.99\textwidth]{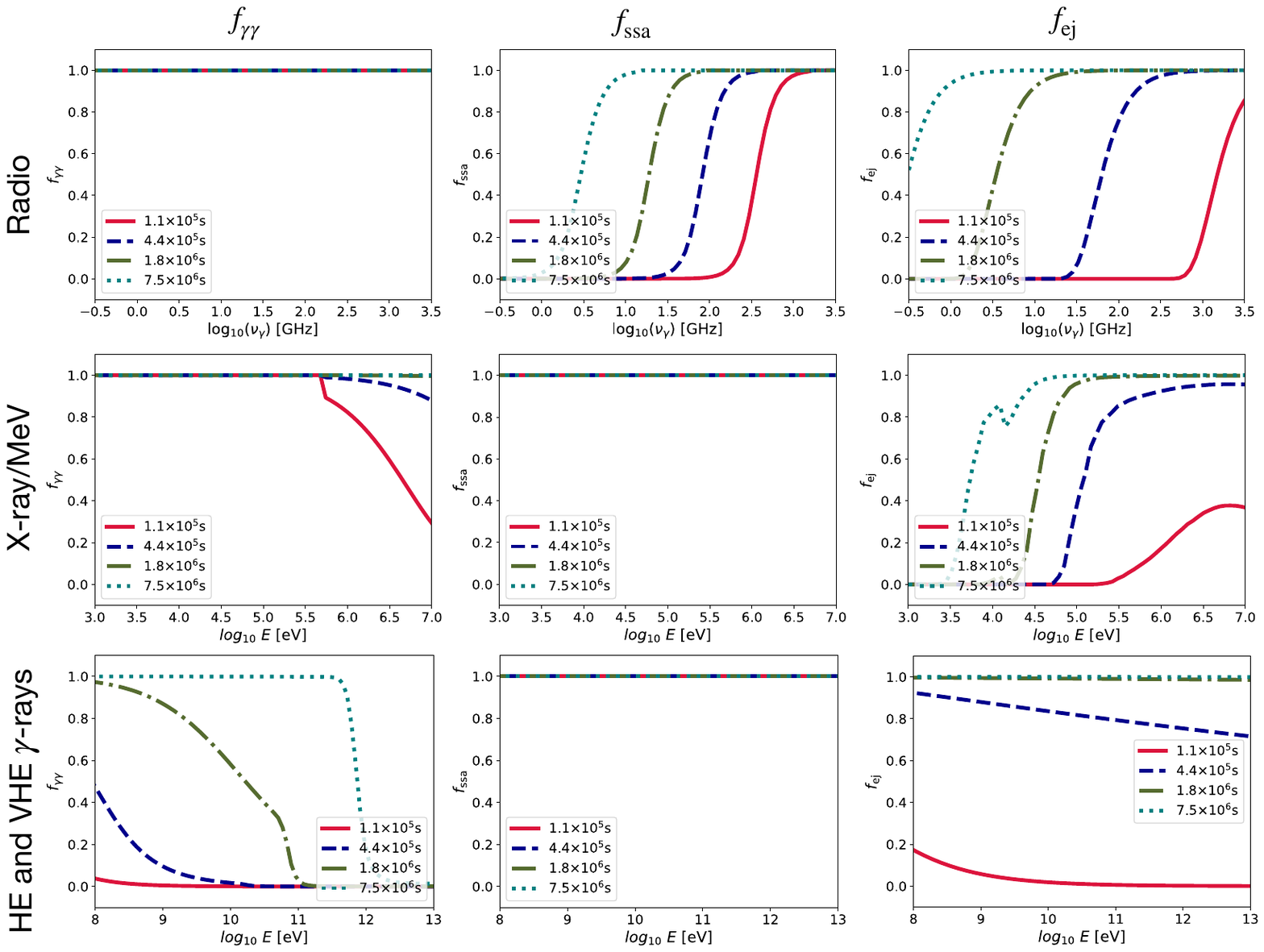}
\caption{\label{fig:attenuations}Attenuation coefficients at different time snaps for $\gamma \gamma$, SSA, and ejecta attenuations denoted by $f_{\gamma \gamma}$, $f_{\rm SSA}$, and $f_{\rm ej}$ respectively. The plots are shown for radio, X-ray/MeV, and high energy (HE) and very high energy (VHE) gamma-rays for the fiducial case. The value of $f=1$ corresponds to $\tau=0$.
}
\end{figure*}
A millisecond pulsar with a strong dipolar magnetic field is formed as a result of a BNS merger. This spinning remnant is long-lived and we refer to it as a \emph{magnetar}. The merger shock and accreting torus post the merger form the outward expanding ejecta. The spindown energy of the magnetar is the main energy reservoir of the system. The rotational energy is converted to magnetar winds consisting of copious amounts $e^+-e^-$ pairs being sourced from the magnetosphere through $B-\gamma$ and $\gamma-\gamma$ processes. This wind interacts with part of the surrounding ejecta from the merger to form a nebular region consisting of $e^+-e^-$ pairs, non-thermal, and thermal photons.

The spindown energy of the magnetar is distributed between the non-thermal, thermal, and magnetic energies. The non-thermal electrons are injected into the nebula by a broken power-law injection spectra (see Section 4.2 in~\citealt{Mukhopadhyay:2024ehs}), $dN/d\gamma_e \sim \gamma_e^{-1.5}$ for $\gamma_m \leq \gamma_e \leq \gamma_{e,
\rm br}$ and $dN/d\gamma_e \sim \gamma_e^{-2.5}$ for $\gamma_{e,
\rm br} < \gamma_e \leq \gamma_M$, where the electron Lorentz factor $\gamma_e = \varepsilon_e/(m_e c^2)$, $\varepsilon_e$ is the comoving energy of the electron, $\gamma_{e,\rm br} = 10^3$ is the break Lorentz factor, $\gamma_m$ and $\gamma_M$ are the minimum and maximum electron Lorentz factors. We set\footnote{Note that the results are insensitive to the choice of $\gamma_m$.} $\gamma_m = 1$ and $\gamma_M$ is determined consistently by balancing the cooling and acceleration timescales for the electrons. The energy distribution of the non-thermal photons is obtained by solving for the steady state transport equation of the electrons, followed by computing the electromagnetic cascades via various processes like synchrotron, inverse Compton (IC), and the subsequent\footnote{The $\gamma\gamma$-absorption is a process in which the low energy soft photons interact with high energy photons to pair-produce ($\gamma + \gamma \rightarrow e^+ + e^-$), such that, $E_\gamma E_{\gamma,\rm soft} \approx (m_ec^2)^2$.} $\gamma \gamma$ processes. The energy density of the non-thermal and thermal photons is obtained by consistently solving for the non-thermal and thermal energy components in the nebular region (see Section 3 in~\citealt{Mukhopadhyay:2024ehs}).

Initially, the system only has non-thermal photons resulting from the EM cascade of the injected non-thermal $e^+-e^-$ pairs. The copious amounts of non-thermal UV- and X-ray photons produced in the nebular region are reprocessed by the effective boundary between the nebula and the ejecta into thermal photons. We quantify this using $\left(1-\mathcal{A}\right)$ which is defined as the \emph{albedo} of the ejecta. In other words, $\mathcal{A}$ gives the fraction of non-thermal photons that escape from the nebular region, while $\left(1 - \mathcal{A}\right)$ gives the fraction of non-thermal photons that are reprocessed into thermal radiation. We compute the albedo $\left(1-\mathcal{A}\right)$ consistently as the system evolves by accounting for the various opacities and relevant processes. The details of this are discussed in Section~\ref{subsec:neb_spec} and Appendix~\ref{app:A}.

The photons in the nebula are attenuated by Synchrotron Self-Absorption (SSA) and $\gamma\gamma$ attenuation processes. Furthermore, the ejecta composed of various heavy elements leads to attenuation across the EM spectrum. The resulting photons along with the Lorentz boost with respect to the observer's frame give the final EM spectrum. We discuss the energy density of the nebular photons, various attenuation processes, and the observable photon spectra in the following section.
\section{Results}
\label{sec:res}
In this section, we present the main results of our work including the nebular photon spectra, emitted photon spectra, and the light curves across various wavelengths. The source is assumed to be at a distance of $100$ Mpc unless otherwise stated. We assume a neutron star remnant with mass $M_* = 2.3 M_\odot$ and radius $R_* = 10$ km. Note that the \emph{fiducial} case with spin period $P_i = 3$ ms, surface equatorial dipole field $B_d = 10^{14}$ G, mass of the ejecta $M_{\rm ej} = 0.03\ M_\odot$, and initial velocity of the ejecta $v_0 = 0.2$c) and the \emph{optimistic} case correspond to the two scenarios considered in~\cite{Mukhopadhyay:2024ehs} (see Table 1 there), where the high-energy neutrino signatures were studied. We mainly discuss the fiducial parameter set unless otherwise noted. The optimistic case has one order-of-magnitude higher spin down energy, leading to higher photon energy density, but the conclusions for the optimistic case will be similar to those for the fiducial case. The results for the optimistic scenario are presented in Appendix~\ref{app:opt_results}.
\subsection{Spectra for nebular photons}
\label{subsec:neb_spec}
We show the distribution of the energy density of non-thermal and thermal photons in the nebular region in Figure~\ref{fig:photonspectra} for two different timesnaps $t_1 = 1.1\times 10^5\ {\rm s}<t_{\rm sd}$ and $t_2 = 1.8\times 10^6\ {\rm s}>t_{\rm sd}$, where the spindown timescale $t_{\rm sd} = 5.6 \times 10^5\ {\rm s}$ for the fiducial case. 
The thermal and non-thermal photons are shown in dashed and solid lines, respectively. For non-thermal photons, the synchrotron and IC components are shown in dot-dashed and dotted lines, respectively.

The thermal photon spectra are assumed to be a blackbody spectrum with temperature given by $T_{\rm th}$ estimated using $T_{\rm th} \sim \left( E_{\rm th}/\left( a_{\rm rad}\ (4/3) \pi R^3 \right) \right)^{1/4}$, where $a_{\rm rad}$ is the radiation constant, $E_{\rm th}$ is the total energy in thermal photons (see Figure 1 in~\citealt{Mukhopadhyay:2024ehs}), and $R$ is the radial distance of the nebula-ejecta boundary from the center of the magnetar. This evaluates to $T_{\rm th}^{t_1} \approx 5.3 \times 10^4$ K and implies $E_\gamma^{\rm th} \sim 2.8 k_B T_{\rm th} \sim 10$ eV (for $E_{\rm th}\approx 7 \times 10^{49}\ {\rm erg},\ R \approx 6.6 \times 10^{14}\ {\rm cm}$). The differential energy density of the thermal photons is thus given by
$u_\nu^{\rm th}=(8\pi h \nu^3/c^3) \left( \exp{(h\nu/(k_B T_{\rm th})) - 1} \right)$, where $\nu$ is the frequency. 

Non-thermal photon spectra are dominantly produced by IC scattering of thermal photons by the electrons injected with a broken-power law distribution, whereas synchrotron emission is relevant only at lower energies. The magnetic energy density in the nebula is estimated to be $u_B \sim \epsilon_B E_{\rm sd}/\left( (4/3) \pi R^3 \right)\sim50\ {\rm erg\ cm}^{-3}$ at $t=t_1$, which is much lower than the thermal photon energy density, $u_\gamma^{\rm th} \sim 4 \times 10^4\ \rm erg\  cm^{-3}$. Since inside the nebula, the $e^+-e^-$ pairs produce non-thermal photons through IC scattering in the fast cooling regime, the peak luminosity of IC scattering will be similar to the spindown luminosity ($L_{\rm sd}$) and peak close to $t_{\rm sd} (= 5.6 \times 10^5\ \rm s)$. This implies the IC energy density peak
\begin{equation}
\nu u_{\nu,{\rm IC}} \sim \frac{L_{\rm nth}}{4 \pi R^2 c} \approx 3 \times 10^4\ {\rm erg}\ {\rm cm}^{-3}\,,
\end{equation}
where the luminosity $L_{\rm nth} \approx L_{\rm sd} \approx 7.13 \times 10^{45}\ {\rm erg\ s}^{-1} \left( B_d/10^{14}\ \rm G \right)^2 \left(P_i/0.003\ \rm s \right)^{-4} \left( 1 + t/t_{\rm sd} \right)^{-2} \sim 8.5 \times 10^{45}\ \rm erg\ s^{-1}$. The peak energy of the non-thermal photon spectra is estimated to be 
\begin{equation}
E_{\gamma,\rm IC} \approx \gamma_{e,\rm br}^2 E_{\gamma}^{\rm th} = 10^7\ {\rm eV} \left(\frac{E_\gamma^{\rm th}}{10\rm~eV}\right)\left(\frac{\gamma_{e,{\rm br}}}{10^3}\right)^2\,,   
\end{equation}
and IC spectrum is soft for $E_\gamma>E_{\gamma,\rm IC}$. The IC photon spectrum has a cutoff due to the Klein-Nishina (KN) effect, which is important if $g=\gamma_e E_{\gamma}^{\rm th}/(m_e c^2)>1$~\citep{Fan:2008cg} is satisfied. This condition is expressed as $\gamma_{e,{\rm KN}}>(m_ec^2)/E_{\gamma}^{\rm th}$. Then, the KN cutoff energy in the IC photon spectrum is given by 
\begin{equation}
E_\gamma^{KN}\approx \gamma_{e,{\rm KN}}^2E_\gamma^{\rm th}\simeq 2.5\times10^{10}\ {\rm eV} \left(\frac{E_\gamma^{\rm th}}{10\rm~eV}\right)^{-1}\,.
\end{equation}

At high energies, $\gamma\gamma$ absorption suppresses the photon spectrum. The thermal photons serve as the target photon field for this process. The photon density is estimated to be $n_\gamma\sim a_{\rm rad} T_{\rm th}^4/E_\gamma^{\rm th}\sim 3\times10^{15}\rm~cm^{-3}$. This leads to a high optical depth for $\gamma\gamma$ absorption (c.f.,~\citealt{Dermer:2009zz}): $\tau_{\gamma\gamma}\approx 0.2\sigma_T n_\gamma R\sim 4\times10^5$ at $t=t_1$. The target photon energy rapidly decreases above $E_\gamma^{\rm th}$, and then we can roughly estimate the $\gamma\gamma$ cutoff energy with the thermal photons using $E_{{\rm cut},\gamma\gamma} \approx (m_e c^2)^2/E_\gamma^{\rm th}\approx E_\gamma^{\rm KN}$. Thus, the cutoff energy by $\gamma\gamma$ process coincides with the KN cutoff.

At lower energies, SSA causes a low-energy cutoff. The SSA optical depth is estimated to be $\tau_{\rm SSA} (\nu) \approx \xi_p e n_e R/(B \gamma_n^5) (\nu/\nu_m)^{-(p+4)/2}$, where $\nu_m \approx \gamma_m^2 e B/(m_e c)$, $n_e$ is the number density of the electrons, $R$ is the radius, $B$ is the comoving magnetic field strength, and $p$ is the spectral index of the electrons post transport which differs from the injection spectrum that we discuss in Section~\ref{sec:model}~\citep[c.f.,][]{Murase:2013kda}. The normalization $\xi_p$ is a function of $p$ (see Equation B4 of~\citealt{Murase:2013kda}). The SSA cutoff frequency is given by $\tau_{\rm ssa}=1$, which provides $E_{\gamma,\rm ssa}\sim0.01$ eV at $t=t_1$ with our set up ($p\simeq2.5$, $R\simeq10^{15}$ cm, $B\simeq30$ G, $\xi_p \sim 18$ and $n_e\simeq10\rm~cm^{-3}$). We should note that we inject electrons with $dN/d\gamma_e\propto \gamma_e^{-1.5}$ in the relevant Lorentz factor, but the spectral index is softened due to cooling. These analytic estimates are roughly consistent with the spectrum shown in Figure~\ref{fig:photonspectra}.

In addition to SSA and $\gamma\gamma$ absorption, the ejecta surrounding the nebula will attenuate the multi-wavelength photons via free-free (radio)~\citep{rybicki_lightman}, bound-bound (IR/Opt/UV), bound-free (UV/X-ray), Compton (MeV), and Bethe-Heitler (GeV-TeV) processes. For the photons of $0.1\ \rm eV<E_\gamma \leq 10$ eV, we set the ejecta opacity to $\kappa_{\rm ej} = 10\ \rm cm^{2}g^{-1}$. For the photons of $10\ \rm eV<E_\gamma \leq 1$ keV, we assume an energy-dependent opacity, $\kappa_{\rm ej}\propto E_\gamma^{1.16}$ in order to smoothly connect the optical opacity to the X-ray one. For the photons of $1\ \rm keV<E_\gamma \leq 20$ MeV, the opacity is computed using the X-ray mass attenuation coefficients\footnote{See Table 3 in~\cite{HubbellSeltzer2004}.}, where the composition ratio of the ejecta consisting of r-process elements is estimated from~\cite{Tanaka:2019iqp,Fujibayashi:2020dvr}. For $E_\gamma>20$ MeV, we take into account Bethe-Heitler (see ~\citealt{Dermer:2009zz}) and Compton attenuations (see \citealt{rybicki_lightman}) based on the composition of the ejecta. 
The optical depth of the ejecta is defined as $\tau_{\rm ej}=\rho_{\rm ej}\kappa_{\rm ej}R$, where $\rho_{\rm ej}=3M_{\rm ej}/(4\pi R^3)$ is the mass density of the ejecta.

We define the attenuation factors as $f_{\gamma\gamma}=(1-e^{-\tau_{\gamma\gamma}})/\tau_{\gamma\gamma}$, $f_{\rm ssa}=(1-e^{-\tau_{\rm ssa}})/\tau_{\rm ssa}$, and $f_{\rm ej}=e^{-\tau_{\rm ej}}$, where $f=1$ corresponds to $\tau=0$.  We plot these in various wave bands (GHz-THz, keV-MeV, GeV-TeV) in Figure~\ref{fig:attenuations}. As discussed above, $\gamma\gamma$ and SSA processes are important only at gamma-ray and radio bands, respectively, whereas ejecta attenuation has influence on all the wavebands at the initial phase. 
As the ejecta and nebula expand, the optical depth decreases and eventually the nebula becomes visible. For radio bands, both SSA and free-free optical depth is lower at higher frequencies, and thus, photons with higher frequencies starts escaping earlier. At late phase of $t\sim7.5\times10^6$ sec, photons of $>10$ GHz can freely escape from the system. For the X-ray/MeV bands, the bound-bound opacity is lower at higher energies, and thus, the MeV photons start escaping earlier than X-ray photons. At late phases, the ejecta becomes transparent to X-rays with energy $E_\gamma>30$ keV. For the high-energy gamma-ray band, $\gamma\gamma$ optical depth is higher for higher energy gamma-rays, and as a result GeV photons start to escape earlier than TeV photons. In the late phase, the ejecta becomes completely transparent, while $\gamma\gamma$ absorption suppresses photons of $\gtrsim1$ TeV. 
\subsection{EM Spectra}
\begin{figure}
\centering
\includegraphics[width=0.47\textwidth]{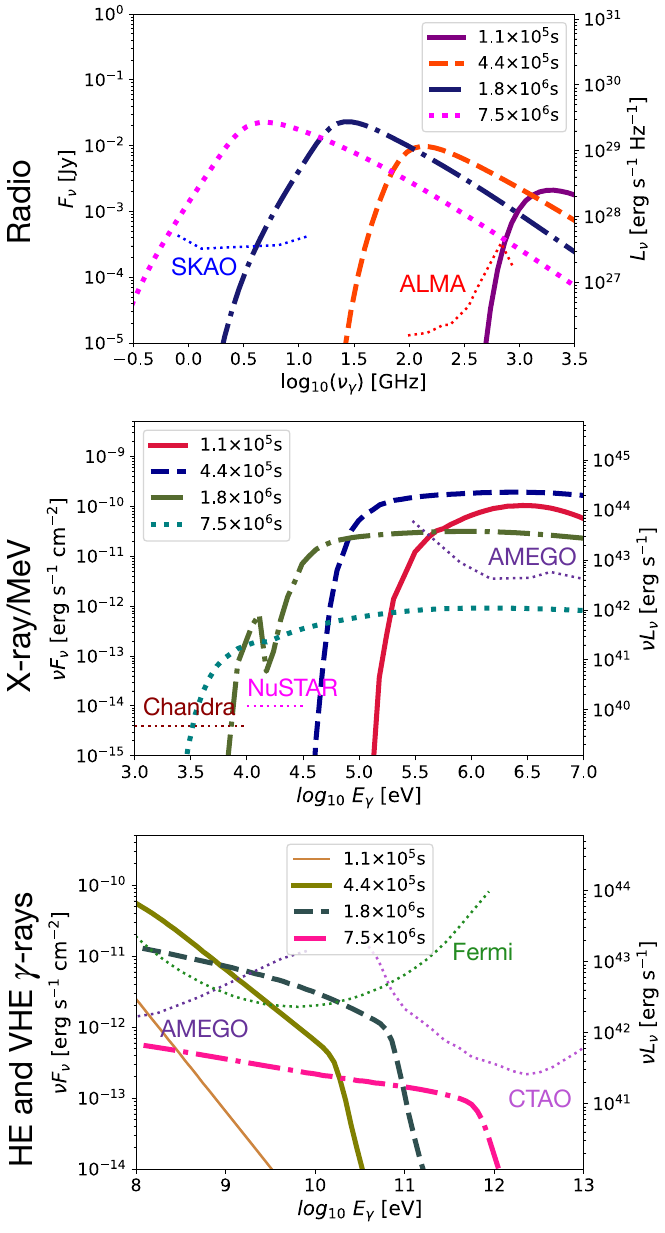}
\caption{\label{fig:emspectra}Observed electromagnetic spectra from a source at $100$ Mpc corresponding to the radio, X-ray, MeV, and gamma-ray bands for the fiducial case at different time snapshots. The sensitivity curves for the relevant detectors in the given EM bands are also shown (see Table~\ref{tab:disthor}). In the plots, $F_\nu$ denotes spectral flux density, $L_\nu$ is the spectral luminosity, and $\nu L_\nu$ is the differential luminosity. Note that the radio spectra (top panel) shows flux density while the middle and bottom panels show flux.
}
\end{figure}
In Figure~\ref{fig:emspectra} we show the observed photon spectra in the mm/radio\footnote{We use \emph{radio} to refer to both mm and radio wavelengths.}, X-ray/MeV, and GeV-TeV gamma-ray bands for different snapshots in time. We also show the sensitivity curves for the relevant detectors (see Table~\ref{tab:disthor}) in various EM bands to evaluate the detection capabilities. In the radio band the spectrum has a cutoff corresponding to the SSA cutoff in the nebula and free-free attenuation in the ejecta which go down in frequency as time increases (see top panel of Figure~\ref{fig:attenuations}). The combination of $f_{\rm SSA}$ and $f_{\rm ej}$ leads to the spectrum having a peak followed by a gradual decline. At initial times, for the fiducial case the peak ($\sim 2$ mJy) occurs at $\mathcal{O}(10^3\ \rm GHz)$, while at very late times the peak ($\sim 20$ mJy) shifts to $\mathcal{O}(1\ \rm GHz)$. Given the sensitivities of SKAO and ALMA, this dictates both detectors can observe the radio wavelengths from such sources for timescales $\gtrsim t_{\rm sd}$.

The attenuation due to heavy elements in the ejecta (see the middle panel if Figure~\ref{fig:attenuations}) completely suppresses the X-ray emission at initial times and thus the X-ray spectra is only significant at very late times ($>10^6$ s). X-ray telescopes like NuSTAR and Chandra have good prospects of X-ray observations from these sources given their typical sensitivities. For MeV photons, we note that attenuation by the ejecta dominates at earlier times ($\sim 1.1 \times 10^5$ s). For $t>t_{\rm sd}$ both the $f_{\gamma \gamma}$ and $f_{\rm ej}$ are close to 1 and hence the MeV photons propagate unattenuated. This explains the approximate plateau shape of the spectra and the decrease in amplitude due to decreasing spindown energy for $t>t_{\rm sd}$. Upcoming gamma-ray detectors like AMEGO have good detection prospects at timescales $t\gtrsim t_{\rm sd}$.

The $\gamma \gamma$ attenuation is most relevant for the high- and very high-energy gamma-rays (see bottom panel of Figure~\ref{fig:attenuations}). The $\gamma \gamma$ cutoff shifts to higher energies with time. Thus, the very high energy gamma-rays are completely suppressed at early times. This is also the case with attenuation in the ejecta due to the Bethe-Heitler and Compton processes. Hence in general, the system is gamma-ray opaque over timescales $t<t_{\rm sd}$. Although the prospects for very high-energy gamma-ray telescopes like CTAO to observe these sources are not optimistic (since the photons at the highest energies are heavily attenuated by the KN and $\gamma \gamma$ attenuation processes), when such observations are available (mostly for remnants $\lesssim 100$ Mpc) CTAO can help distinguish such remnants from the pulsar-wind-nebulae in our own Galaxy, which are very bright in the VHE gamma-ray range. Lastly, $100$ MeV to $10$ GeV gamma-rays can be observed using AMEGO or Fermi respectively.
\subsection{Light curves}
\begin{figure}
\centering
\includegraphics[width=0.48\textwidth]{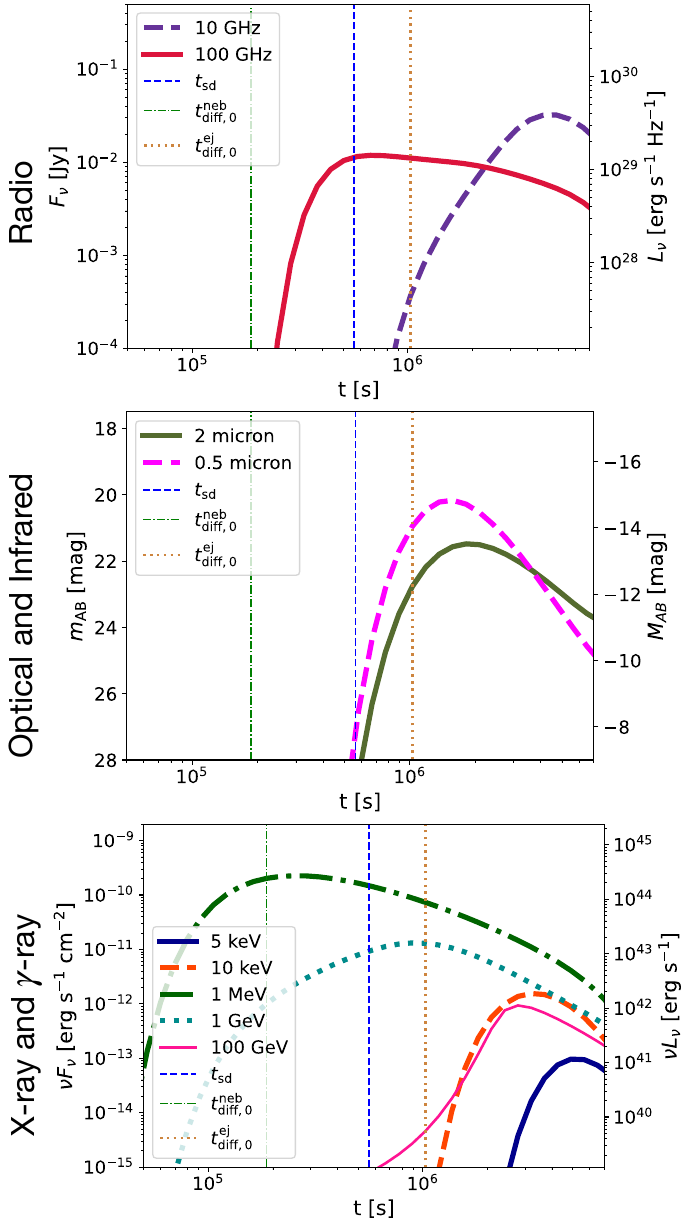}
\caption{\label{fig:lc}Light curves from a source at $100$ Mpc corresponding to the radio, optical, infrared, X-ray and gamma-ray bands for the fiducial case. The characteristic spindown ($t_{\rm sd}$), nebular diffusion ($t_{\rm diff,0}^{\rm neb}$), and ejecta diffusion  ($t_{\rm diff,0}^{\rm ej}$) timescales are also shown. The representative frequencies (for radio), wavelengths (for optical and infrared), and energies (for X-ray and $\gamma-$ rays) are motivated by relevant bands from observational telescopes shown in Table~\ref{tab:disthor} and Figures~\ref{fig:emspectra} and~\ref{fig:disthor}. In the plots, $F_\nu$ denotes spectral flux density, $L_\nu$ is the spectral luminosity, the apparent and absolute magnitudes are shown by $m_{AB}$ and $M_{AB}$ respectively, and $\nu L_\nu$ is the differential luminosity. Note that $t$ is the time elapsed from the instance of the BNS merger.
}
\end{figure}
The light curves corresponding to the radio, optical, infrared, X-ray and gamma-ray bands for the fiducial case is shown in Figure~\ref{fig:lc}. The light curves can be effectively understood from Figure~\ref{fig:emspectra}. For the radio band at $10$ GHz, the photon spectra is non-negligible only at late times ($\sim  10^6$ s), before which the SSA cutoff suppresses the emission and thus the peak occurs $\sim 4 \times 10^6$ s. The high frequency radio emission at $100$ GHz peaks around the spindown timescale $t_{\rm sd}$ corresponding to the SSA cut-off when $\tau_{\rm SSA} = 1$. This is followed by a slow decline, which is primarily associated with a decrease in the photon energy density in the nebula.

The optical and infrared bands with wavelengths $0.5\ \mu m$ and $2\ \mu m$ respectively peak at late times ($\sim 2 \times 10^6$ s). The most notable feature here is at the peak the AB magnitude is roughly 20 (21.5) for the optical (infrared) wavelengths. This is comparable to the kilonova brightness observed for GW170817 \citep{J-GEM:2017tyx,Tanaka:2017qxj}, where for GW170817 the AB magnitudes would be 19.4 (20) scaled from $\sim 40$ Mpc to $100$ Mpc in the optical (infrared) wavelengths. Thus for long-lived magnetar driven systems, one can expect a late time \emph{re-brightening} much later than the typical kilonova timescales. This can potentially provide evidence for a surviving remnant.

The X-ray bands corresponding to $5$ and $10$ keV are strongly attenuated until late times and peak around timescales $>2 \times 10^6$ s. This can be attributed to the heavy r-process elements in the ejecta that only becomes transparent to X-rays at late times (see middle panel of Figure~\ref{fig:attenuations}). The time when the ejecta becomes transparent to X-ray photons can be analytically estimated by setting $\tau_{\rm ej}=1$, which leads to 
\begin{align}
t_{\rm pk} &\approx \sqrt{\frac{3 M_{\rm ej} \kappa}{4 \pi v_0^2}}\\
&\simeq 61.3\ {\rm days}\ \left(\frac{M_{\rm ej}}{0.03M_\odot}\right)^{1/2}\left(\frac{v_0}{0.2c}\right)^{-1}\left(\frac{\kappa_{\rm ej}}{70}\right)^{1/2}\nonumber,
\end{align}
where we use $\kappa_{\rm ej}$ for 10 keV photons. For 1 MeV photons, $\kappa_{\rm ej}\sim 0.06\rm~cm^{-2}~s^{-1}$, with which we obtain $t_{\rm pk}\simeq1.5\times10^5$ s. These estimates agree with what we find in Figure~\ref{fig:lc}.

The $1$ GeV high energy gamma-rays peak and attain a maxima on timescales comparable to the ejecta diffusion timescale prior to which they suffer heavy attenuation due to $\gamma \gamma$ interactions. At very late times the ejecta becomes transparent to these gamma-rays. In this case, the peak will roughly be given by time $t=\rm max \left[ \tau_{\gamma \gamma}(t) \sim 1, \tau_{\rm ej}(t) \sim 1 \right]$. The ejecta opacity in the 1 GeV band, $\kappa_{\rm ej}^{\rm GeV} \sim 0.095\ \rm cm^{2}g^{-1}$ which gives $t_{\tau_{\rm ej} \sim 1} \sim 2 \times 10^5$ s when $\tau_{\rm ej} \sim 1$, while $t_{\tau_{\gamma \gamma \sim 1}}$ can be obtained by setting $\tau_{\gamma \gamma} \sim 1$. Since the latter is larger we see that the GeV gamma-rays peak around $t_{\tau_{\gamma \gamma} \sim 1}$. Finally, the very high energy $100$ GeV gamma-rays are heavily suppressed at timescales $t<t_{\rm sd}$. Post the spindown timescale, the ejecta becomes transparent to the very high energy gamma-rays when the $\gamma \gamma$ cutoff shifts to higher energies and the attenuations ($f_{\gamma \gamma}$ and $f_{\rm ej}$) become negligible.
\section{Prospects for Detection}
\label{sec:det}
\subsection{Detection horizon for EM telescopes}
\label{subsec:dethor_EM}
\begin{figure}
\centering
\includegraphics[width=0.48\textwidth]{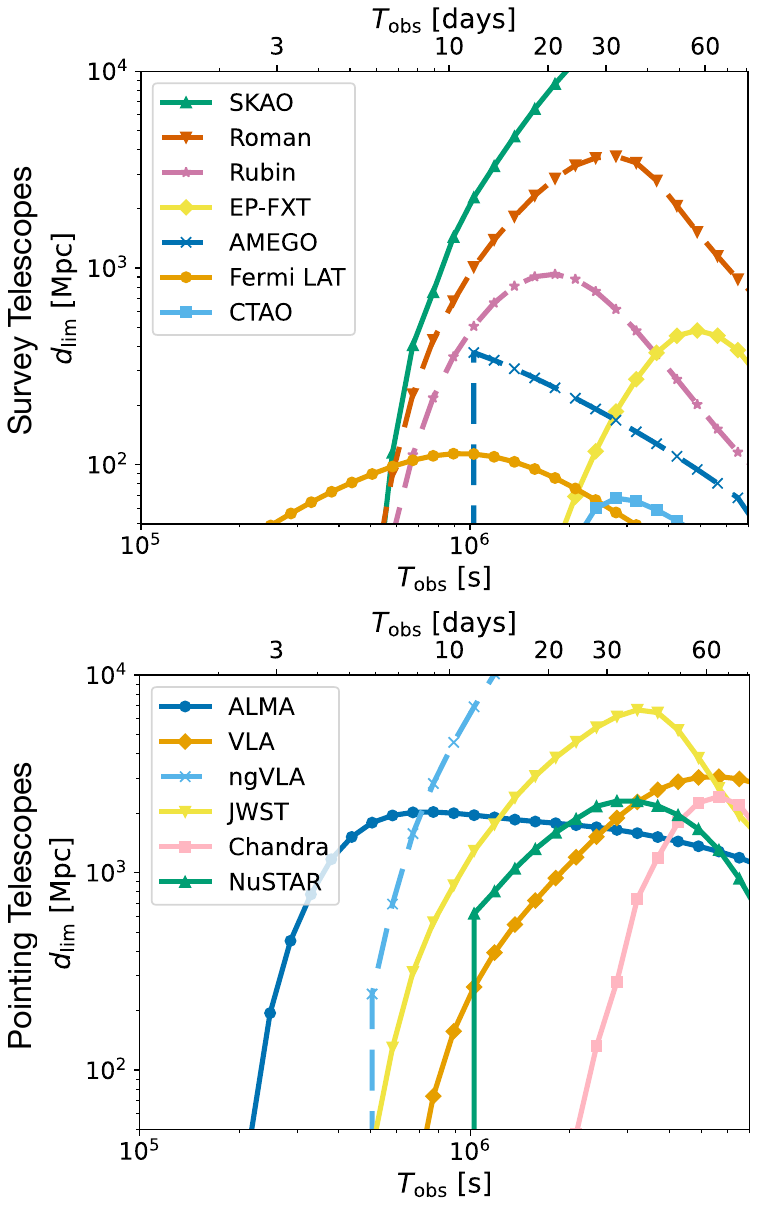}
\caption{\label{fig:disthor}Distance horizons ($d_{\rm lim}$ see Equation~\ref{eq:dlim}) for various survey and pointing telescopes for the fiducial scenario,. The upcoming telescopes are shown in dashed lines. See Table~\ref{tab:disthor} for details about the telescopes. Note that for EP-FXT, AMEGO, and NuSTAR, the sudden jump is a result of the long integration time $\Delta T_{\rm int} = 10^6$ s. Note that $T_{\rm obs}$ is the time elapsed from the GW detection of the BNS merger.
}
\end{figure}
\begin{table*}[t]
\centering
\scriptsize
\setlength{\tabcolsep}{2pt}
\caption{Details of various telescopes used in Figure~\ref{fig:disthor}: sensitivity $F_{\rm sens}$, relevant frequency $\nu$/energy $E_\gamma$/wavelength $\lambda_\gamma$, integration time $\Delta T_{\rm int}$, maximum horizon distance $d_{\rm lim}^{\rm max}$, time post the merger for maximum horizon distance $t(d_{\rm lim}^{\rm max})$. The results for the optimistic case are shown within square brackets.\\
\it{Additional details: }\emph{SKAO}: Square Kilometre Array Observatory (\href{https://www.skao.int/en/science-users/ska-tools/493/ska-sensitivity-calculators}{https://www.skao.int}), \emph{Roman}: Nancy Grace Roman Space Telescope (\href{https://www.stsci.edu/files/live/sites/www/files/home/roman/observatory/_documents/roman-capabilities-table.pdf}{https://www.stsci.edu}), \emph{Rubin}: Vera C. Rubin Observatory - Large Synoptic Survey Telescope (LSST) (\href{https://smtn-002.lsst.io}{https://smtn-002.lsst.io}), \emph{EP-FXT}: Einstein Probe - Follow-up X-ray Telescope~\citep{Zhang:2021pad},  \emph{AMEGO}: All-sky Medium Energy Gamma-ray Observatory~\citep{Kierans:2020otl}, \emph{Fermi LAT}: Fermi Gamma-ray Space Telescope Large Area Telescope (LAT) (\href{https://fermi.gsfc.nasa.gov/ssc/data/analysis/documentation/Cicerone/Cicerone_Introduction/LAT_overview.html}{https://fermi.gsfc.nasa.gov}), \emph{CTAO}: Cherenkov Telescope Array Observatory (\href{https://www.ctao.org/for-scientists/performance/}{https://www.ctao.org}), \emph{ALMA}: Atacama Large Millimeter Array (\href{https://almascience.nao.ac.jp/proposing/sensitivity-calculator}{https://almascience.nao.ac.jp}), \emph{VLA}: Karl G. Jansky Very Large Array (\href{https://obs.vla.nrao.edu/ect/}{https://obs.vla.nrao.edu}), \emph{ngVLA}: next-generation VLA~\citep{selina2018generationlargearraytechnical}, \emph{JWST}: James Webb Space Telescope (\href{https://www.stsci.edu/jwst/science-planning/proposal-planning-toolbox/sensitivity-and-saturation-limits}{https://www.stsci.edu/jwst}), \emph{Chandra}: Chandra X-ray Observatory using High-Resolution Camera (HRC) (\href{https://cxc.harvard.edu/cdo/about_chandra/}{https://cxc.harvard.edu}), \emph{NuSTAR}: Nuclear Spectroscopic Telescope Array (\href{https://heasarc.gsfc.nasa.gov/docs/nustar/nustar_tech_desc.html}{https://heasarc.gsfc.nasa.gov}).}
\begin{tabular}{p{2.5cm} p{3.5cm} p{2.0cm} p{2.0cm} p{2.0cm} p{2.0cm}}
\hline
\hline
Telescope & $F_{\rm sens}$ & $\nu/E_\gamma/\lambda_\gamma$ & $\Delta T_{\rm int}$ & $d_{\rm lim}^{\rm max}$ & $t(d_{\rm lim}^{\rm max})$ \\
&  & & (s) & (Gpc) & (days) \\
\hline
\multicolumn{6}{c}{\textbf{Survey Telescopes}} \\
\hline
SKAO & $1.2 \times 10^{-6}\ {\rm Jy}$ & $12.5$\ \rm GHz & $3.6 \times 10^3$ & $24.0$ [$62.7$] & $75$ [$115$] \\
Roman & $4.6 \times 10^{-14}\ \rm erg\ cm^{-2}\ s^{-1}$ & $1.5\ \mu$m & $3.6 \times 10^3$ & $3.70$ [$18.3$] & $32$ [$87$] \\
Rubin & $2.7 \times 10^{-15}\ \rm erg\ cm^{-2}\ s^{-1}$ & $0.5\ \mu$m & $3.0 \times 10^1$ & $0.93$ [$3.00$] & $21$ [$43$] \\
EP-FXT & $7.0 \times 10^{-15}\ \rm erg\ cm^{-2}\ s^{-1}$ & $5$\ \rm keV  & $1.0 \times 10^6$ & $0.48$ [$2.75$] & $57$ [$100$] \\
AMEGO & $1.6 \times 10^{-12}\ \rm erg\ cm^{-2}\ s^{-1}$ & $100$\ \rm MeV & $1.0 \times 10^6$ & $0.37$ [$0.99$] & $12$ [$14$] \\
Fermi LAT & $1.0 \times 10^{-11}\ \rm erg\ cm^{-2}\ s^{-1}$ & $1$\ \rm GeV & $3.0 \times 10^4$ & $0.11$ [$0.20$] & $10$ [$28$] \\
CTAO & $2.0 \times 10^{-12}\ \rm erg\ cm^{-2}\ s^{-1}$ & $100$\ \rm GeV & $3.0 \times 10^4$ & $0.07$ [$0.14$] & $32$ [$65$] \\
\hline
\multicolumn{6}{c}{\textbf{Pointing Telescopes}} \\
\hline
ALMA & $3.4 \times 10^{-5}\ {\rm Jy}$ & $100$\ \rm GHz & $3.6 \times 10^3$ & $2.03$ [$4.95$] & $9$ [$16$] \\
VLA  & $5.0 \times 10^{-5}\ {\rm Jy}$ & $10$\ \rm GHz & $3.6 \times 10^3$ & $3.06$ [$7.26$] & $65$ [$115$] \\
ngVLA & $1.9 \times 10^{-7}\ {\rm Jy}$ & $16$\ \rm GHz & $3.6 \times 10^3$ & $65.6$ [$184$] & $49$ [$115$] \\
JWST & $8.2 \times 10^{-9}\ {\rm Jy}$ & $2\ \mu$m & $1.0 \times 10^4$ & $6.67$ [$40.5$] & $37$ [$133$] \\
Chandra & $9.0 \times 10^{-16}\ \rm erg\ cm^{-2}\ s^{-1}$ & $5$\ \rm keV & $3.0 \times 10^5$ & $2.43$ [$13.7$] & $65$ [$133$] \\
NuSTAR & $1.0 \times 10^{-14}\ \rm erg\ cm^{-2}\ s^{-1}$ & $20$\ \rm keV & $1.0 \times 10^6$ & $2.30$ [$10.7$] & $32$ [$57$] \\
\hline
\end{tabular}
\label{tab:disthor}
\end{table*}
In this section, we discuss the detection prospects associated with long-lived magnetar remnants from BNS mergers at current and upcoming EM telescopes and outline the observational strategy based on the source distance. The redshift-dependent limiting distance or horizon distance $d_{\rm lim}$ given an observed flux $F_\nu^{\rm obs}$ at a given observed frequency $\nu_{\rm obs}$, luminosity distance $d_L$, and observation time $T_{\rm obs}$ post the merger can be defined as,
\begin{align}
\label{eq:dlim}
F_\nu^{\rm obs} (\nu_{\rm obs}, T_{\rm obs}, d_L) &= \frac{(1+z) L_{\nu_{\rm src}}(T_{\rm src})}{4 \pi d_L^2}\,,\\
\int_{T_{\rm obs}}^{T_{\rm obs}+\Delta T_{\rm int}}\hspace{-0.8cm}dt_{\rm obs}\ F_\nu^{\rm obs}  (\nu_{\rm obs}, &T_{\rm obs}, d_L = d_{\rm lim}) = F_{\rm sens} (\nu_{\rm obs},\Delta T_{\rm int})\,\nonumber,
\end{align}
where $F_{\rm sens}$ is the sensitivity of the corresponding EM telescope at a given frequency $\nu$ given an integration time or exposure time $\Delta T_{\rm int}$ in the observer frame. The observer frequency and time, $\nu_{\rm obs} = \nu_{\rm src}/(1+z)$ and $T_{\rm obs} = (1+z)T_{\rm src}$ respectively, where $z$ is the redshift. The horizon distance ($d_{\rm lim}$) is obtained by numerically solving Equation~\ref{eq:dlim}.

We consider various current and upcoming EM telescopes and in particular distinguish between the survey and pointing telescopes. While the former can quickly scan large portions of the sky (even multiple times a day for telescopes like Fermi LAT and AMEGO) to find a source, the latter help in focusing on and characterizing the sources with high precision. The survey telescopes we consider include SKAO in the radio band, Roman and Rubin in the optical/infrared band, EP-FXT in the X-ray band, AMEGO in the MeV band, Fermi LAT and CTAO in the high-energy and very-high energy gamma-ray bands\footnote{EP's Wide-Field X-ray Telescope does not have sufficient sensitivity to detect these objects. Moreover, we do not discuss High Altitude Water Cherenkov (HAWC) Observatory and Large High Altitude Air Shower Observatory (LHAASO) as gamma-rays above 1 TeV is severely attenuated.}. While for the pointing telescopes we have ALMA, VLA, and ngVLA in the radio band, Chandra and NuSTAR in the X-ray band. In Figure~\ref{fig:disthor} we show the horizon distances for both the survey and pointing telescopes. The details of $\Delta T_{\rm int}$, $\nu$, and $F_{\rm sens}$ are given in Table~\ref{tab:disthor}. In Figure \ref{fig:disthor} and Table \ref{tab:disthor}, we do not take into account the time lag between the Target of Opportunity (ToO) trigger time and follow-up observations. For example, ALMA and NuSTAR have 9-day and 2-day time lag between the trigger time and actual observation, respectively. However, these timescale is much shorter than the duration of the transient, and thus, we can ignore the time lag to obtain the values of $d_{\rm lim}^{\rm max}$ and $t(d_{\rm lim}^{\rm max})$.

For a BNS merger within $100$ Mpc, MeV and high-energy gamma-rays (for distances $\mathcal{O}(50)$ Mpc even very high-energy gamma-rays can be observed using CTAO\footnote{We ignore extra-galactic background light (EBL) absorption in this case.}) can be identified using AMEGO and Fermi LAT approximately a week post the merger respectively. For such a fortunate scenario, EM signatures from the remnant will be detectable across all wavebands including optical. This will result in a clear identification of the host galaxy and its redshift using optical spectroscopic observations. Since the gamma-ray distance horizon is close ($d_{\rm lim}^{\rm max} \sim 0.11$ Gpc), one can estimate the number of such long-lived magnetar remnants that can be expected in existing telescopes like Fermi-LAT. Assuming the fiducial ($z=0$) BNS merger~\citep{KAGRA:2021duu} is $\rho_0 \approx 300\ {\rm Gpc}^{-3}{\rm yr}^{-1}$ and the number of BNS mergers forming such stable remnants $f_{\rm mag} \sim 10\%$, we have the number of stable remnants ($N_{\rm BNS}^{\rm stable}$) that could have been seen by Fermi LAT\footnote{However, Fermi does not issue alerts (although the data is public) which makes the situation for searches non-trivial. Dedicated searches have not been performed on the Fermi data to look for these sources.} within an operation timescale $T_{\rm op} \approx 15$ years to be $N_{\rm BNS}^{\rm stable} \approx f_{\rm mag}\rho_0 (4/3)\pi \left(d_{\rm lim}^{\rm max}\right)^3 \sim 2 - 3$.

For BNS mergers occurring within $\mathcal{O}(1)$ Gpc, survey telescopes like SKAO, Roman, Rubin, EP-FXT can discover radio, optical, infrared, and X-ray counterparts within timescales of a week-months. However, for these cases gamma-ray observations are challenging. The survey telescope observations can be followed up by more sensitive pointing telescopes, like ALMA, VLA, JWST, Chandra, and NuSTAR in the respective wavebands. Moreover, observations from Chandra or JWST, will help in host galaxy identification and provide information about redshift when spectroscopic follow-up observations are available. In addition, ALMA and VLA can yield precise localizations due to their sub arc-second localization capabilities (dependent on their configuration), which helps in uniquely associating the merger event with its EM counterpart, as discussed in Section~\ref{subsec:multi_wavelength_id}.

Finally, in the case of BNS mergers several Gpc away, only radio and possibly infrared observations are likely to be made using survey telescopes like SKAO and Roman on timescales of $\mathcal{O}(10)$ days after the merger. In such cases, precise follow-up observations can be performed using JWST yielding crucial host galaxy and redshift information. However, it would be challenging to characterize the unique features of remnants.  This is because the radio transient detected as a result cannot be uniquely associated with the detected GW signature from the BNS merger (see Section~\ref{subsec:multi_wavelength_id}).
\subsection{Multi-wavelength signatures to identify magnetar-powered transients}
\label{subsec:multi_wavelength_id}
In this section, we qualitatively discuss and highlight the importance of classifying a transient as a long-lived magnetar remnant resulting from a BNS merger. The major challenge in this regard is the following. In general, GW detection of a BNS merger with the current detectors will have not so high SNRs, leading to huge uncertainty in localization\footnote{This is assuming an absence of a short GRB or when the GRB is off-axis.} $\sim \mathcal{O}(10 - 100\ \rm deg^2)$. As discussed in Section~\ref{subsec:dethor_EM} the survey telescopes with their large detection horizons will indeed discover a large number of transients including supernovae, GRBs, TDEs. In such a scenario, how can one uniquely associate the detection to a magnetar-powered transient resulting from a BNS merger? 

\begin{table*}[ht]
\centering
\caption{Summary of multi-wavelength signatures to identify magnetar-powered transients}
\begin{tabular}{p{0.08\linewidth} p{0.3\linewidth} p{0.4\linewidth} p{0.1\linewidth}}
\hline
\hline
$d_L$ [Mpc] & Primary Backgrounds & Strategy & Unique Id.\\
\hline
$\lesssim 100$ & GRBs from collapsars and mergers & $\gamma-$ray observations from MeV (in AMEGO) to very high energies (in Fermi LAT and possibly CTAO). Long timescales $\mathcal{O}(10^5 {\rm s} - 10^6 \rm s)$ & Yes \\[4pt]
\hline
$\lesssim 1000$ & Supernovae (optical and radio), Galactic transients (X-ray), Kilonova (optical/near infrared) 
& (a) Plateau in the radio (100 GHz) band, (b) Rebrightening in the optical and infrared band, (c) Late time rise of hard and soft X-rays. 
& Yes \\[4pt]
\hline
$> 1000$ & Transients across all wavelengths & Radio observations, Roman & Challenging\\[4pt]
\hline
\end{tabular}
\label{tab:id_summ}
\end{table*}

The above issue is resolved by leveraging the multi-wavelength electromagnetic spectrum along with their temporal evolution as discussed in Section~\ref{sec:res}, which serve as characteristic of magnetar-powered transients discussed in this work. The GW observation of the BNS merger event will provide a distance estimate. Thus, we structure our discussion by the distance to the source. Note that all timescales discussed below are with respect to the GW detection of the BNS merger event. We discuss some specific classes of transients that can serve as backgrounds for EM observations and highlight properties of EM emission that can effectively help distinguish emission from magnetar-powered remnants from BNS mergers. Our results are succinctly summarized in Table~\ref{tab:id_summ}.

For mergers within 100 Mpc, the \emph{smoking gun} evidence for a stable magnetar remnant would be the observation of gamma-rays from MeV (in AMEGO) to very high energies (in Fermi LAT and possibly CTAO). This is because the possible backgrounds for this scenario would be GRBs (both from collapsars or compact object mergers)~\citep{2022Galax..10...74G,2022Galax..10....7N}. However, on-axis GRBs have very different timescales associated with them and hence the light curves between a GRB and a magnetar-powered remnant can be easily distinguishable. The gamma-ray emission from GRBs, although brighter initially, eventually fades out. Whereas from Figure~\ref{fig:lc} it is evident that in our scenario, the gamma-rays increase with time and peak $\sim$ week-month post the merger. For an off-axis GRB afterglow (in the standard forward shock scenario) as in GW170817, the rising timescale is indeed similar ($\sim$ months), in which case the late time rise of very high-energy gamma-rays will help distinguish the magnetar remnants from mergers\footnote{GeV-TeV gamma-ray lightcurves are expected to be similar to X-ray lightcurves~\citep{Miceli:2022efx}.}. Observations using Fermi LAT in the 1 GeV gamma-ray band along with CTAO in the 100 GeV band continue for a long time and is much brighter than afterglow emissions, which serve as key to distinguish these from off-axis GRBs. Furthermore, the MeV photons peak earlier than the gamma-rays and their observation using AMEGO can be followed up by observations using Fermi LAT to conclusively establish the transient to be powered by a magnetar central engine.

A less fortunate (but more probable) scenario is GW signatures from BNS mergers within $\sim 1$ Gpc. This would be within the detection horizon for survey telescopes in the radio (SKAO), optical (Rubin), infrared (Roman), and X-ray (EP-FXT) bands. In the optical and radio band, supernovae will act as the main background, while in the X-ray band various galactic transients like stellar flares, novae, accretion flows in low mass X-ray binaries, serve as the main background. For example, in the optical band, Rubin will see $\mathcal{O}(10^2)$ supernovae a day, making it impossible to distinguish magnetar-powered transients from various other transients based on the optical light curve. This degeneracy can be broken by combining the observations across the four wavelengths and carefully looking at the light curve information. As seen from Figure~\ref{fig:lc}, the classic feature of magnetar-powered remnants is that the light curves peak on timescales of a week-months which can then enable more dedicated observations on a set of selected transients using pointing telescopes.

In such a scenario, one can use high frequency ($100$ GHz) radio observations from ALMA that have a characteristic plateau following a rise, to uniquely identify emission from magnetar-powered BNS remnants. Moreover, observations in the optical band using JWST, where one would see the light curve brightening around a week after the merger, can not only help in disentangling the merger remnant, but would also help in predicting the redshift. Similarly in case of X-rays, the late rise in the light curve serves as a distinguishable feature and can help with unique associations. Thus, when combining EM light curves across wavelengths, a magnetar-powered transient embedded in r-process–enriched ejecta would exhibit a distinctive sequence: a peak in the $100$ GHz band first, followed by peaks in the optical/IR band, and finally peaks in the soft ($5$ keV) and hard ($10$ keV) X-ray bands. It is interesting to note that for supernovae the X-ray peak is weaker and the associated timescale of the peak is also different. Furthermore, some TDEs show signs of late time X-ray emissions~\citep{Bower:2012ga}, but in such transients the radio band light curve has a slower rise time~\citep{Cendes:2023rrc}.

Perhaps one of the most sought after follow-up observations post a GW detection of a BNS merger is a kilonova in the optical/near infrared bands. Magnetar remnant surrounded by the enriched r-process ejecta can still be disentangled from kilonova observations using the duration of the kilonova emission which is longer for the former. Besides, strong radio and X-ray emissions are typically not expected from kilonovae. Thus radio and X-ray observations in addition with the optical/infrared observations can clearly disentangle the scenario discussed in this work from kilonova signatures.

Finally for BNS mergers at distances $> 1$ Gpc, radio survey telescopes will need to use the temporal information from the light curves to distinguish magnetar-powered remnants from various other transients. However, this is challenging with just radio observations. In case Roman jointly observes a delayed rising of the light curve, dedicated observations can be implemented using JWST for some selected transients, which can ultimately lead to unique identification.

It is important to note that optical spectroscopy is frequently used to firmly classify an optical transient. In our scenario, optical signals from the nebula should be a featureless continuum because of the synchrotron nature. These featureless nebula emissions need to pass through the r-process enriched ejecta, possibly causing some absorption/emission features by r-process elements as seen in GW170817~\citep{Waxman:2017sqv,Domoto:2022cqp,Vieira:2022tnm,Vieira:2023rbc,Hotokezaka:2023aiq} and GRB 230307~\citep{JWST:2023jqa}. The prediction of features demands non-equilibrium radiative transfer calculation with external non-thermal photon fields, which we leave for future work. 
\section{Conclusions}
\label{sec:disc}
Long-lived remnants from BNS mergers can lead to enhanced electromagnetic emission across all wavelengths. In this work, we studied a system consisting of a millisecond pulsar with a large dipolar magnetic field ($\sim 10^{13}\rm G - 10^{15}$G) (termed as a magnetar) resulting from a BNS merger, surrounded by an expanding nebula and kilonova ejecta\footnote{See Figure 1 in~\cite{Mukhopadhyay:2024ehs} for a schematic figure.}. The basic idea is that the magnetar spins down while losing energy through magnetic dipole radiation. A dominant fraction of this rotational energy is converted to magnetar winds consisting of $e^+-e^-$ pairs. These winds interact with the surrounding material ejected at the time of merger (ejecta), forming a nebula consisting of $e^+e^-$ pairs, thermal and non-thermal photons. We show the nebular photon spectra in Figure~\ref{fig:photonspectra}. 

We consistently account for attenuation both in the nebular and ejecta regions as shown in Figure~\ref{fig:attenuations}. While the former is caused primarily by $\gamma \gamma$ and SSA processes, the latter is the sum of the optical, infrared, X-ray (due to heavy r-process elements in the ejecta), Compton, and Bethe-Heitler attenuations along with free-free absorption across the whole spectrum. The resulting EM spectra and lightcurves are shown in Figures~\ref{fig:emspectra} and~\ref{fig:lc} for a typical source at $100$ Mpc. The parameter set chosen for our results are similar to GW170817, while the results for a neutrino-optimistic parameter set are presented in Appendix~\ref{app:opt_results}. We note that the EM emissions although initially suppressed due to heavy attenuation, peak at timescales of week-month post the merger, and present very optimistic detection prospects at timescales of $\mathcal{O}(1 - 10)$ days post the merger.

Besides exploring inclination-independent (unlike in GRBs) non-thermal multi-wavelength EM spectra, we also quantify the capabilities of current and upcoming, survey and follow up telescopes by computing their detection horizons in Figure~\ref{fig:disthor}. Furthermore, we outline (in Section~\ref{subsec:multi_wavelength_id}) features in the multi-wavelength spectra that can be used to uniquely identify magnetar remnants from BNS mergers, given the large localization area that will be associated with GW observations of BNS mergers. In particular, we can disentangle the magnetar-powered transients discussed here from supernovae using high frequency radio observations (for example using ALMA), along with observations of MeV photons, followed by high-energy gamma-ray observations (using Fermi LAT). Besides this, optical spectroscopy can be considered for unique associations. 

We also note that our estimates for $d_{\rm lim}$ (see Equation~\ref{eq:dlim}) are robust in significance since we average (integrate) over a time-interval of $\Delta T_{\rm int}$ above the threshold sensitivity of the telescope. This means we have a time-averaged flux and not just fluctuations above the telescope sensitivities. The limits do not change as long as the telescopes have an opportunity to visit the source multiple times. Moreover, given the long timescales, a dedicated observation cycle ToO is not needed for survey telescopes like Rubin-LSST, Fermi LAT, AMEGO, and SKAO, for which the usual survey mode should be sufficient for performing significant observations of such sources. However telescopes like CTAO, EP-FXT, and Roman would need to conduct dedicated searches and hence need a ToO to identify these sources. Long-term multi-wavelength EM follow-up observations for future GW events would enable us to unravel the properties of BNS merger remnants in the near future.
\begin{acknowledgements}
We thank Brian D. Metzger, David Radice for comments. M.\,M. is particularly grateful to Eduardo Mario Guti\'errez for many useful discussions.
M.\,M. wishes to thank the Astronomical Institute at Tohoku University, for their hospitality where a major part of this work was completed. M.M. is also grateful to CERN and the Institute of Theoretical Physics of the Jagiellonian University for hospitality and partial support during the last stages of this work.
M.\,M. is supported by NSF Grant No. AST-2108466. M.\,M. also acknowledges support from the Institute for Gravitation and the Cosmos (IGC) Postdoctoral Fellowship. S.S.K. acknowledges the support by KAKENHI No. 22K14028, No. 21H04487, No. 23H04899, and the Tohoku Initiative for Fostering Global Researchers for Interdisciplinary Sciences (TI-FRIS) of MEXT’s Strategic Professional Development Program for Young Researchers.
\end{acknowledgements}
\bibliography{refs}{}
\bibliographystyle{aasjournalv7}
\appendix

\section{Appendix A: On the time evolution of albedo of the ejecta $\left(1-\mathcal{A}\right)$} 
\label{app:A}
\begin{figure*}
\centering
\includegraphics[width=0.48\textwidth]{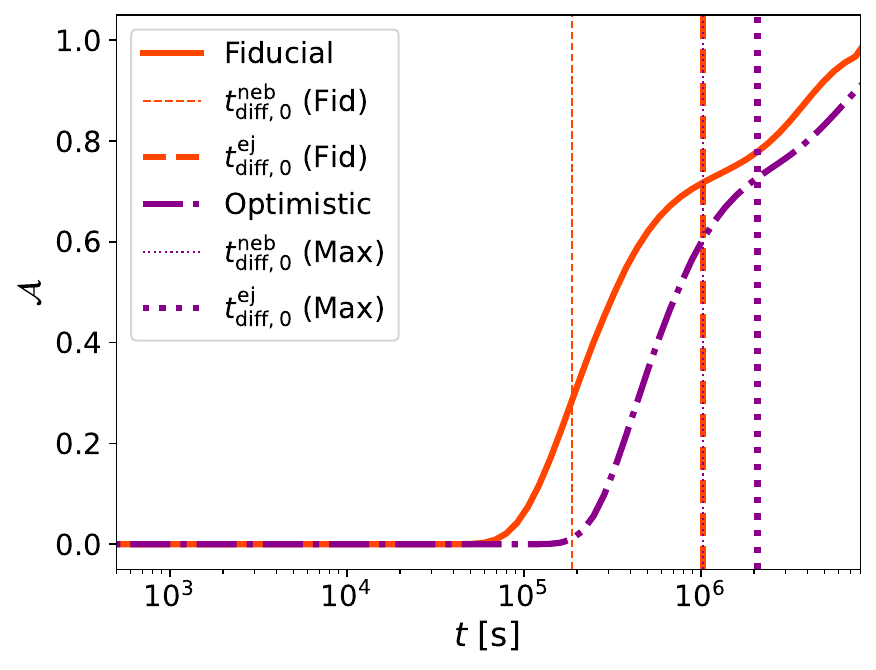}
\caption{\label{fig:curlya} Time evolution of $\mathcal{A}$ for the fiducial (solid redorange) and the optimistic (dot-dashed purple) scenarios.
}
\end{figure*}
We define $\mathcal{A}$ in the following way
\be
\mathcal{A} = \frac{\int_{\varepsilon_{\rm min}}^{\varepsilon_{\rm max}}\ d\varepsilon\ \varepsilon L_\varepsilon^{\rm obs}}{\int_{\varepsilon_{\rm min}}^{\varepsilon_{\rm max}}\ d\varepsilon\ \varepsilon L_\varepsilon^{\rm intrinsic}}\,,
\ee
where the observed luminosity is defined as $L_\varepsilon^{\rm obs} = L_\varepsilon^{\rm intrinsic} - L_\varepsilon^{\rm attenuated}$ and $\varepsilon$ is the energy of the photons in the comoving frame. The intrinsic luminosity $L_\varepsilon^{\rm intrinsic}$ is defined as the luminosity of the non-thermal and thermal photons that are injected into the ejecta post $\gamma \gamma$ absorption and SSA in the nebular region. Finally the attenuated luminosity $L_\varepsilon^{\rm attenuated}$ quantifies the attenuation of the non-thermal and thermal photons in the ejecta (see also Section~\ref{subsec:neb_spec}) due to free-free absorption (radio), a Planck-mean bound-bound opacity (infrared/optical/ultraviolet), a wave-length dependent bound-free and Compton scattering opacity (X-ray to soft gamma-rays), and Bethe-Heitler opacity (high-energy gamma-rays).

We show the time-evolution of $\mathcal{A}$ in Figure~\ref{fig:curlya}. Complete attenuation is characterized by $\mathcal{A} = 0$, while $\mathcal{A} = 1$ denotes complete transmission. We note both for the fiducial and optimistic cases, $\mathcal{A} \sim 0$ for $t\lesssim 10^5$ s. The relevant timescales across which $\mathcal{A}$ varies are given by the diffusion timescales across the nebular and ejecta regions. At very late times $t\gtrsim 10^7$ s $\mathcal{A} \rightarrow 1$ signifying the transparency of the ejecta. Note that $\mathcal{A}$ is the fraction of non-thermal photons that escape from the system, while a fraction $(1-\mathcal{A})$ of the non-thermal photons are reprocessed and act as the source of thermal photons in the nebular region.
\section{Results for the optimistic case} 
\label{app:opt_results}
\begin{figure}
\centering
\includegraphics[width=0.48\textwidth]{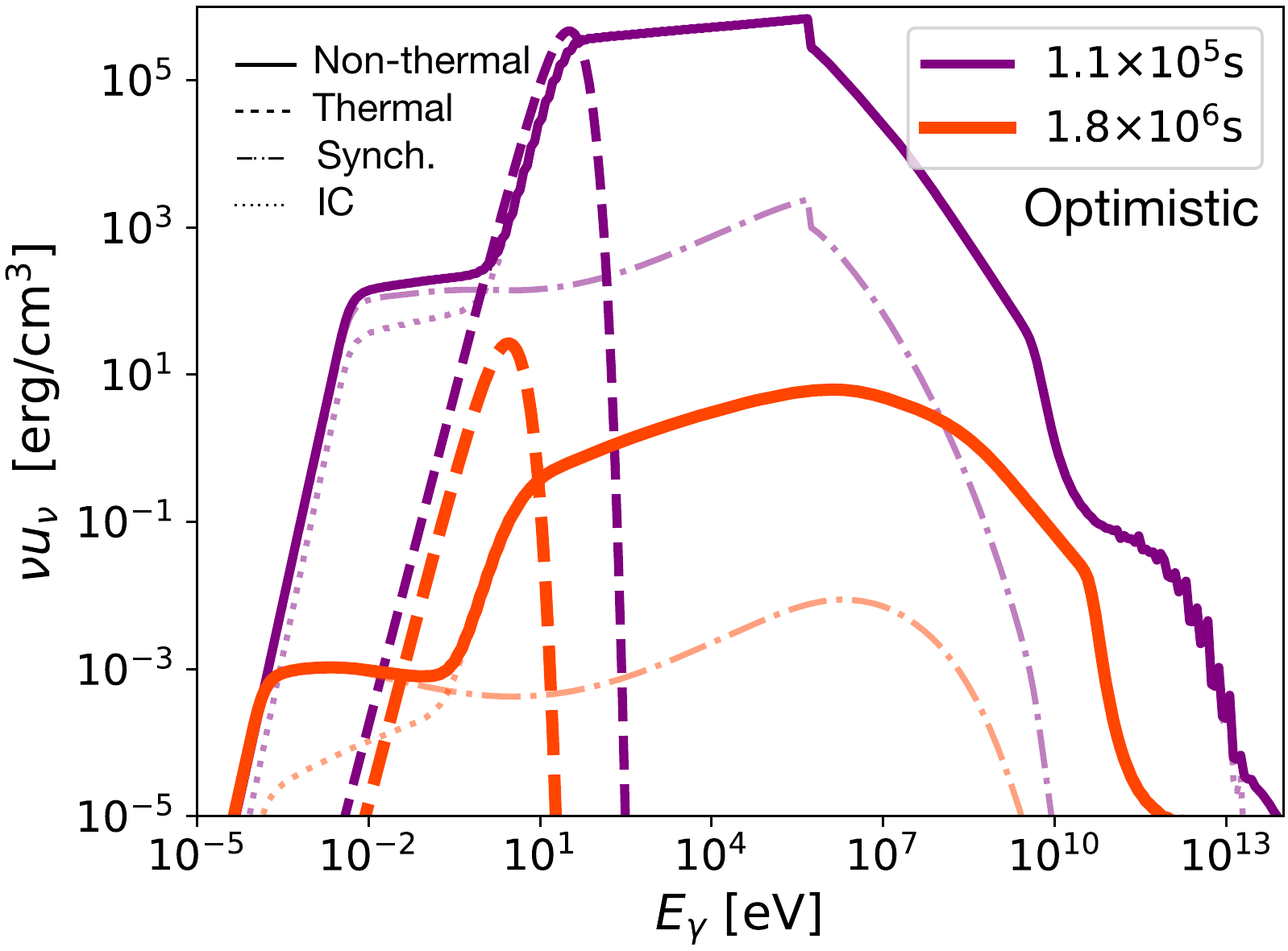}
\caption{\label{fig:opt_photonspectra}Same as Figure.~\ref{fig:photonspectra} but for the optimistic case.
}
\end{figure}
\begin{figure*}
\centering
\includegraphics[width=0.99\textwidth]{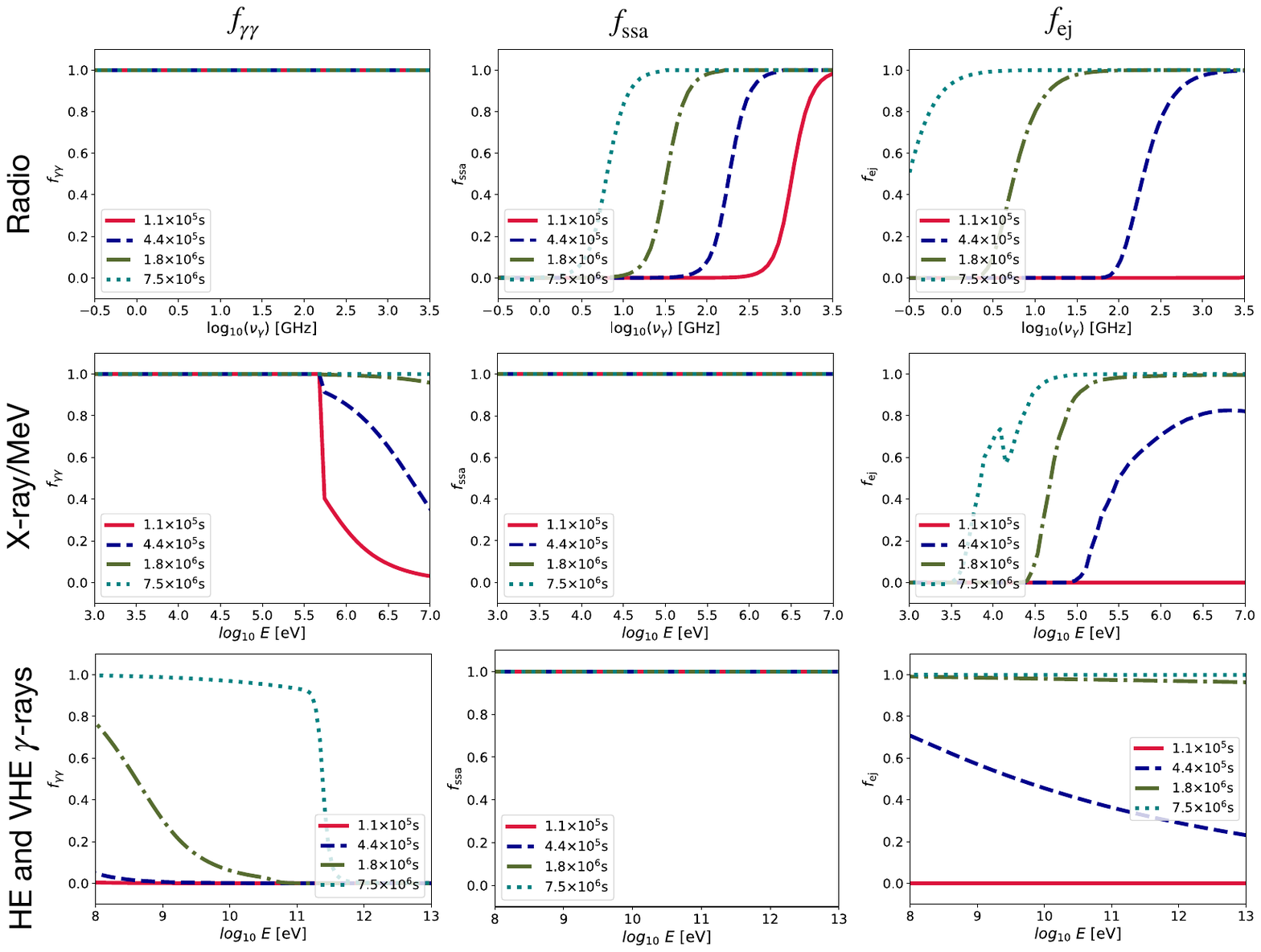}
\caption{\label{fig:opt_attenuations}Same as Figure.~\ref{fig:attenuations} but for the optimistic case.
}
\end{figure*}
\begin{figure*}
\centering
\includegraphics[width=0.4\textwidth]{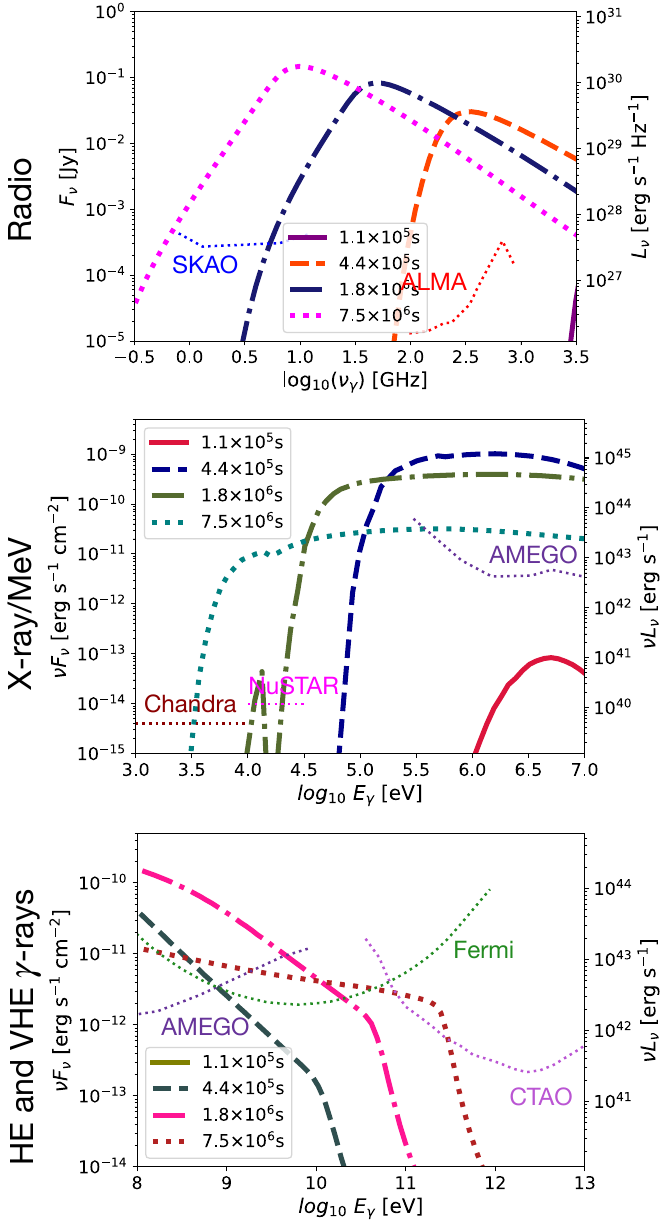}
\includegraphics[width=0.4\textwidth]{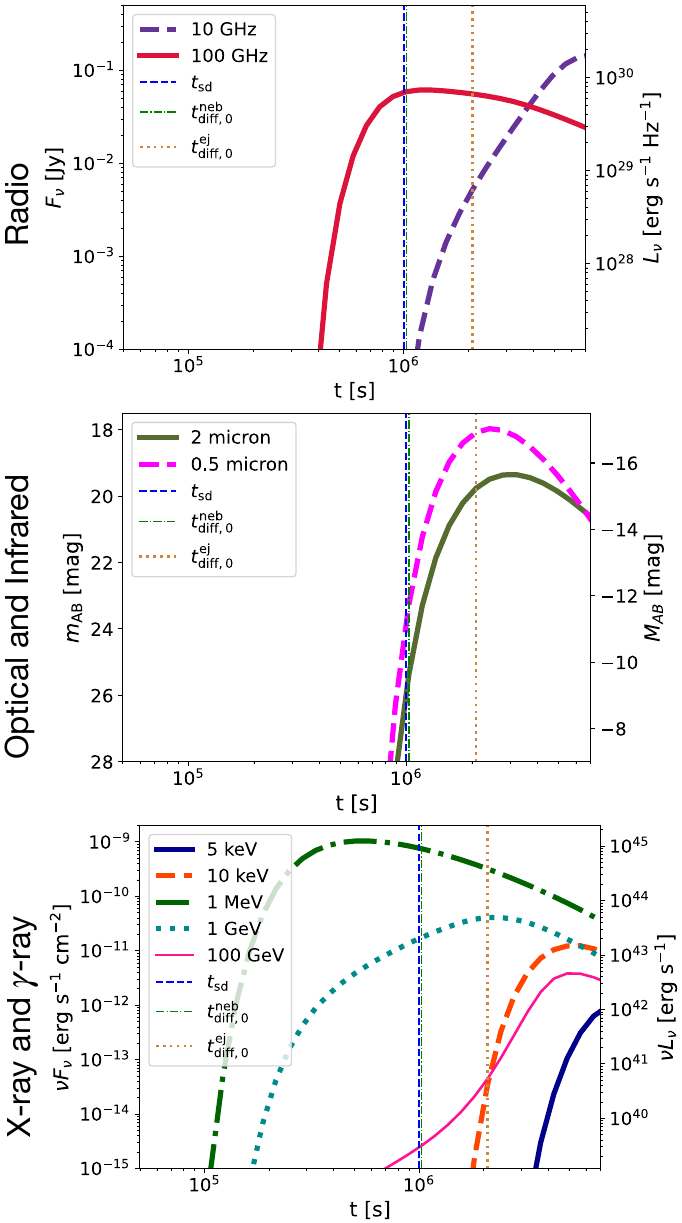}
\caption{\label{fig:opt_emspectra_lc} \emph{Left: }Same as Figure.~\ref{fig:emspectra} but for the optimistic case. \emph{Right: }Same as Figure~\ref{fig:lc} but for the optimistic case.
}
\end{figure*}
\begin{figure*}
\centering
\includegraphics[width=0.98\textwidth]{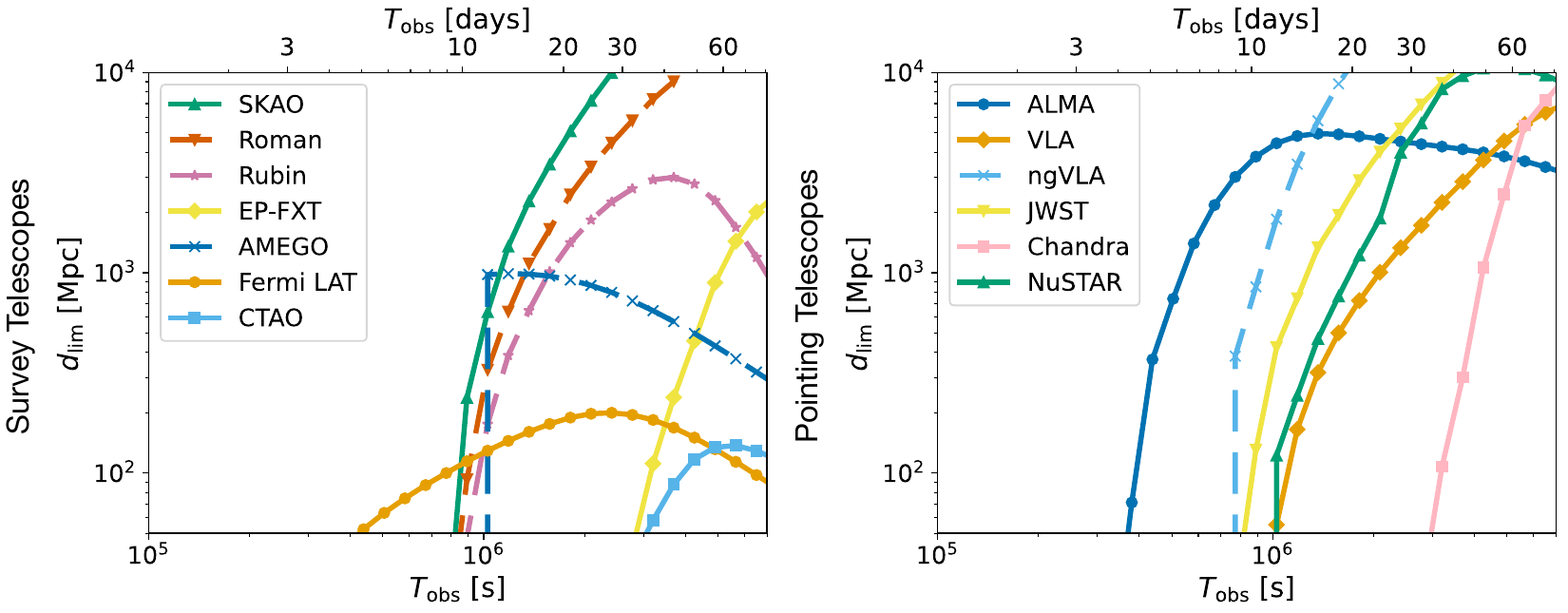}
\caption{\label{fig:opt_disthor}Same as Figure~\ref{fig:disthor} but for the optimistic case.
}
\end{figure*}
In this Appendix we present the results for the optimistic parameter set ($P_i = 1$ ms, $B_d = 2.5 \times 10^{13}$ G, $M_{\rm ej} = 0.1\ M_\odot$, $v_0 = 0.1$c). The spindown timescale $t_{\rm sd} = 1.0 \times 10^6\ {\rm s}$ for the optimistic case. Compared to the fiducial case, it has one order of magnitude higher spindown energy and a spindown timescale $t_{\rm sd} \sim 10^6$ s. Thus the conclusions for this parameter set is similar to the fiducial case except that the emissions peak around the $\gtrsim 10^6$ s and are higher in magnitude than the fiducial case.

The energy density of the non-thermal and thermal photons are shown in Figure~\ref{fig:opt_photonspectra}. The various attenuation coefficients are shown in Figure~\ref{fig:opt_attenuations}. The observed electromagnetic spectra (\emph{left}) and the light curves (\emph{right}) corresponding to various bands - radio, optical/infrared, X-ray, MeV, gamma-rays - are shown in Figure~\ref{fig:opt_emspectra_lc}. Finally, the distance horizons for various survey and pointing telescopes are shown in Figure~\ref{fig:opt_disthor}.
\end{document}